\documentclass[reprint,
nofootinbib,
amssymb,
prd,
unsortedaddress,
]{revtex4-1}

\usepackage[intlimits]{amsmath}
\usepackage{accents}

\usepackage{graphicx}% Include figure files
\usepackage{dcolumn}% Align table columns on decimal point
\usepackage{bm}% bold math
\usepackage[english]{babel}

\usepackage{tikz}
\usetikzlibrary{arrows.meta}
\usepackage[compat=1.1.0]{tikz-feynman}

\usepackage{verbatim}

\usepackage[colorlinks=true, linkcolor = blue,
urlcolor  = blue,
citecolor = blue]{hyperref}

\newcommand*{\punkt}[1]{%
	\accentset{\mbox{\large \bfseries .}}{#1}}

\begin{document}

\title{Avoiding parameter fine-tuning in mass varying neutrino models of DE?}

\author{Michael~Maziashvili}
\email{maziashvili@iliauni.edu.ge} 
\author{Vakhtang~Tsintsabadze}

\affiliation{\vspace{0.2cm} {School of Natural Sciences and Medicine, Ilia State University, \\ 3/5 Cholokashvili Ave., Tbilisi 0162, Georgia}}

% \date{\today}% It is always \today, today,
             %  but any date may be explicitly specified

\begin{abstract}  
Coupled models of quintessence are usually introduced for avoiding or mitigating the parameter fine-tuning problem. At the same time, the coupled models should avoid the fine-tuning problem related to the initial conditions as quintessence models do. One more attractive feature of coupled models might be the explanation of the timescale at which the coincidence of DE and matter energy densities occur and the understanding of reason why after this the DE takes over. And finally, all these nice features should be unaffected by the quantum corrections of the potential. Having in mind these remarks, we shall focus our discussion on mass varying neutrino model of DE with inverse power-law potential to see how naturally does it work.

%\begin{description}

%\item[PACS numbers]

%\end{description}
\end{abstract}

%\pacs{Valid PACS appear here}% PACS, the Physics and Astronomy
                             % Classification Scheme.
%\keywords{Suggested keywords}%Use showkeys class option if keyword
                              %display desired
\maketitle
%\tableofcontents

\section{Introduction}
\label{intro}

One of the widely considered possibilities for the dark energy (DE) with time-varying equation of state parameter is the quintessence \cite{Weiss:1987xa, Wetterich:1987fm, Peebles:1987ek, Ratra:1987rm, Zlatev:1998tr, Steinhardt:1999nw}. One of the main motivations of such an approach is to understand the appearance of a particular time-scale at which the DE and matter energy densities become of the same order of magnitude and also the clarification of the question what makes DE to become predominant after this coincidence. As far as we are talking about the uncoupled model of quintessence, the situation looks as follows. Since for uncoupled fluids the energy-momentum tensors are separately conserved, that is, 

\begin{eqnarray}
&&\dot{\rho}_j = -3H\rho_j (1+\omega_j) \Rightarrow \nonumber \\&&  \rho_j(a) = \rho_j(1)\exp\left(3\int_a^1\left[1+\omega_j(a)\right]\mathrm{d}\ln a\right) ~, \nonumber 
\end{eqnarray} it follows that

\begin{eqnarray}
\frac{\rho_\phi(a)}{\rho_M(a)} = \frac{\rho_i(1)}{\rho_j(1)} \exp\left(3\int_a^1\left[\omega_\phi(\tilde{a})-\omega_M(\tilde{a})\right]\mathrm{d}\ln \tilde{a}\right) ~, ~~
\end{eqnarray} where the subscript $M$ stands for the dominating matter component, which maybe either relativistic or nonrelativistic, and $a$ is set to unity at the time when DE takes over. Thus, the DE model requires that $\rho_\phi < \rho_M \Leftrightarrow \omega_\phi < \omega_M$ for $a< 1$ and $\rho_\phi \gtrsim \rho_M$ for $a>1$. That is, we need to have a proper synchronization between $\rho_\phi$ and $\rho_M$ the model to work. On the other hand, one should take care that the presence of $\rho_\phi$ is harmless for early time cosmology. Namely, for nucleosynthesis and structure formation.

Assuming $\dot{\omega}_\phi \approx 0$ for the late time evolution, from equations

\begin{eqnarray}
	 U(\phi) = \rho_\phi(1-\omega_\phi)/2 ~,  ~~  \label{conteq} \dot{\rho}_\phi = -\, 3H\rho_\phi(1+\omega_\phi)  ~, \nonumber
\end{eqnarray} one obtains that (here $\dot{\phi}$ is assumed to be positive)

\begin{eqnarray}\label{slope}
\frac{U'}{U} = - \, \frac{3H(1+\omega_\phi)}{\dot{\phi}} = - \, \sqrt{\frac{24\pi (1+\omega_\phi)}{M_P^2\Omega_\phi}} ~.
\end{eqnarray} Taking the time derivative of Eq.\eqref{slope}, it follows that

\begin{eqnarray}\label{trackingp}
&&\frac{U''U - (U')^2 }{(U')^2} \, \frac{U'}{U} \equiv (\Gamma - 1) \frac{U'}{U}  =  - \frac{1}{2} \frac{\dot{\Omega}_\phi}{\Omega_\phi} ~ \Rightarrow  \nonumber \\&&  (\Gamma - 1) \frac{3H(1+\omega_\phi)}{\dot{\phi}} =  \frac{1}{2} \frac{\dot{\Omega}_\phi}{\Omega_\phi} ~. 
\end{eqnarray} From Eq.\eqref{trackingp} it is clear that for successful quintessence models the relation $\Gamma \equiv U''U/ (U')^2  \geq 1$ should hold. One well known example satisfying this criterion is the Ratra-Peebles model based on inverse-power law potential \cite{Ratra:1987rm, Peebles:1987ek}

\begin{eqnarray}\label{potential}
U(\phi) = V \left(\frac{M_P}{\phi}\right)^\alpha ~. 
\end{eqnarray} It is an example of runaway model indicating that the field rolls down the potential monotonically - going to "infinity". The DE in this model is a slow roll phenomenon. Namely, the slow roll parameters 

\begin{eqnarray}\label{slow}
	\frac{M_P^2}{16\pi} \left(\frac{U'}{U}\right)^2  = \frac{M_P^2}{16\pi} \left(\frac{\alpha}{\phi}\right)^2 ~, ~   \frac{M_P^2}{8\pi} \, \frac{U''}{U} = \frac{M_P^2}{8\pi} \, \frac{\alpha(1+\alpha)}{\phi^2} ~, ~~~~
\end{eqnarray} manifest that the field enters slow roll regime for $\phi \gtrsim M_P$ and, rougly speaking, the condition $\phi \simeq M_P$ marks the onset of DE. Therefore, in view of the potential \eqref{potential}, the parameter $V$ should be taken to be of the same order of magnitude as $\rho_{DE}^0$. Besides this parameter fine-tuning, an important point worth noting is that as we are dealing with large-filed model of quintessence, $\phi \gtrsim M_P$, the (one-loop) quantum corrections to the potential may turn out to be unacceptably large \cite{Kolda:1998wq, Doran:2002bc}. One may hope to avoid these problems by introducing a coupling of $\phi$ with the cosmic neutrino background (CNB) in a manner suggested in \cite{Fardon:2003eh}. The idea behind this coupling is to introduce the energy scale associated to $\rho_{DE}^0$ from particle physics sector. Namely, this energy scale is quite close to the neutrino mass scale. Besides, one may hope that the time-scale at which neutrinos enter the non-relativistic regime could be dynamically related to the DE activation time. We shall consider the coupling of the form

\begin{eqnarray}\label{coupling}
 m_\nu(\phi) = \mu_\nu \left( \frac{\phi}{\mu_\nu} \right)^\beta ~. 
\end{eqnarray} Let us note that the parameters $\beta$ in Eq.\eqref{coupling} and $\alpha$ in Eq.\eqref{potential} are both understood to be of order unity. In Eq.\eqref{coupling} we have introduced one more dimensional parameter of the model - the neutrino mass scale: $\mu_\nu$. The main purpose of the present discussion is to see how naturally does this coupled model work. Under the naturalness of model we understand its capability of providing self-tuning of parameters in course of cosmological evolution and the dynamical introduction of particular time-scales. We shall first discuss some basic features of the mass varying neutrino model for DE. It is done in Section \ref{general}. Next, in Section \ref{attractor}, we consider the attractor nature of dynamics for the model under consideration. The linearized analysis near the fixed point fails in our case because one of the Eigenwerten is zero and one needs to exploit center manifold technique. The Section \ref{approximate} is devoted to the discussion of an approximate solution obtained by minimizing of the effective potential. This solution is quite accurate so long as the model remains in the small field regime. In the next Section the model is examined against the one-loop radiative corrections. As the model is non-renormalizable, extra care is needed in finding a reasonable way for evaluating the size of one-loop quantum corrections. The discussion is summarized in Section \ref{discussion}.

Before proceeding let us mention a few papers about the mass varying neutrino models in which various interesting aspects are worked out \cite{Gu:2003er, Peccei:2004sz, Bi:2004ns, Brookfield:2005bz, Honda:2005tw, Fardon:2005wc, Rosenfeld:2005pw, delaMacorra:2007xkb, Peccei:2006qx, Gu:2007mi, Chitov:2009ph, Ando:2009ts, Bamba:2008jq, MohseniSadjadi:2017pne, Sadjadi:2018xqo, Mandal:2019kkv, Esteban:2021ozz, Anari:2022tsl}.

Let us note that throughout of this paper we shall use natural units: $c = \hbar = 1$, in which $G_N^{-1/2}=M_P \approx 1.2\times 10^{28}$\,eV and, in addition, the subscript an superscript zero will be used to denote present values of various quantities.

\section{Some general features of the model}
\label{general}

The model is based on the action functional \cite{Wetterich:2007kr, Amendola:2007yx}

\begin{eqnarray}&&
\int\mathrm{d}^4x\,\sqrt{-g} \left( \frac{g^{\alpha\beta}\partial_\alpha\phi\partial_\beta\phi}{2} \,-\, U(\phi) \,-\,  \frac{M_P^2R}{16\pi } \,+ \right. \nonumber \\&& \left. \frac{i}{2}\left[\bar{\psi}_\nu\gamma^\alpha(x)\mathfrak{D}_\alpha\psi_\nu - \big(\mathfrak{D}_\alpha\bar{\psi}_\nu\big)\gamma^\alpha(x)\psi_\nu  \right] - m_\nu(\phi)\bar{\psi}_\nu\psi_\nu  \right) ~, \nonumber
\end{eqnarray} in spatially flat FLRW metric 

\begin{eqnarray}\label{FLRW}
\mathrm{ds}^2 = \mathrm{d}t^2 - a^2(t)\mathrm{d}\mathbf{x}^2 ~. 
\end{eqnarray} The fermion field here is assumed to stand for the cosmic neutrinos, which decouple from the heat/thermal bath of cosmic soup at about $1.5$ MeV and form after that CNB. From the fermion sector, the variation with respect to $\phi$ results in $-m'_\nu\bar{\psi}_\nu\psi_\nu$. This term can be expressed in terms of the Spur of neutrino stress-energy tensor

\begin{eqnarray}\label{fermioncontribution}
-m'_\nu\bar{\psi}_\nu\psi_\nu = \frac{m'_\nu}{m_\nu} T^\alpha_{~\alpha} = \frac{m'_\nu}{m_\nu} (\rho_\nu - 3p_\nu) ~. 
\end{eqnarray} Next one assumes a finite-temperature expression for $\rho_\nu - 3p_\nu$ in order to account for the properties of CNB. Summarizing, the equations of motion read

\begin{eqnarray}
&& \dot{\rho}_\nu + 3H(\rho_\nu +p_\nu) =  \frac{\mathrm{d}\ln m_\nu}{\mathrm{d}\phi}(\rho_\nu - 3p_\nu)\dot{\phi} ~, \label{continuity} \\ && 
\ddot{\phi} +3H\dot{\phi} +U'(\phi) = -\frac{\mathrm{d}\ln m_\nu}{\mathrm{d}\phi} (\rho_\nu - 3p_\nu) ~, \label{eqofmot} \\ && H^2 = \frac{8\pi}{3M_P^2}(\rho_\nu+\rho_\phi +\rho_r+\rho_m) \label{Friedmann}~.
\end{eqnarray} Here $\rho_r$ and $\rho_m$ denote radiation and matter densities, respectively and   

\begin{eqnarray}
&&\rho_\phi = \frac{\dot{\phi}^2}{2} +U(\phi) ~.  \nonumber
\end{eqnarray} As for the neutrinos, we assume that they are free streaming,

\begin{eqnarray}
&& \rho_\nu = \frac{\mathsf{g}}{a^3}\int\frac{\mathrm{d}^3k}{(2\pi)^3}\, \frac{\varepsilon_\nu(\mathbf{k})}{\mathrm{e}^{k/aT_\nu} + 1} \equiv \mathsf{g} \int\frac{\mathrm{d}^3q}{(2\pi)^3}\, \frac{\varepsilon_\nu(\mathbf{q})}{\mathrm{e}^{q/T_\nu} + 1}  ~,  \nonumber \\&& p_\nu  =  \frac{\mathsf{g}}{3a^5}\int\frac{\mathrm{d}^3k}{(2\pi)^3}\frac{k^2}{\varepsilon_\nu(\mathbf{k})\Big(\mathrm{e}^{k/aT_\nu} + 1\Big)} \equiv \nonumber \\ && ~~~~~~~~~~~ \frac{\mathsf{g}}{3}\int\frac{\mathrm{d}^3q}{(2\pi)^3}\frac{q^2}{\varepsilon_\nu(\mathbf{q})\Big(\mathrm{e}^{q/T_\nu} + 1\Big)}  ~, \nonumber \\ && \varepsilon_\nu(\mathbf{k}) = \sqrt{\frac{\mathbf{k}^2}{a^2}+m_\nu^2(\phi)} \equiv \sqrt{\mathbf{q}^2+m_\nu^2(\phi)} ~,  \nonumber
\end{eqnarray} where the factor $\mathsf{g}$ counts two helicity states per flavor and is thus equal to  

\begin{eqnarray}
\mathsf{g} = 2\times \text{number of neutrino species} ~. \nonumber 
\end{eqnarray} One can immediately check that the Eq.\eqref{continuity} is "automatically" satisfied by these expressions of $\rho_\nu$ and $p_\nu$. Now it is straightforward to verify that the quantity $ (\rho_\nu - 3p_\nu)m'_\nu/m_\nu $ can be conveniently written as $\partial\rho_\nu/\partial\phi$. Namely, one finds 

\begin{eqnarray}
\frac{\partial\rho_\nu}{\partial\phi} = 	\mathsf{g} m_\nu m'_\nu \int\frac{\mathrm{d}^3q}{(2\pi)^3}\, \frac{1}{\varepsilon_\nu(\mathbf{q})\left[\mathrm{e}^{q/T_\nu} + 1\right]} ~, \nonumber  \\   \rho_\nu - 3p_\nu = 	\mathsf{g}m^2_\nu \int\frac{\mathrm{d}^3q}{(2\pi)^3}\, \frac{1}{\varepsilon_\nu(\mathbf{q})\left[\mathrm{e}^{q/T_\nu} + 1\right]} ~,  \nonumber 
\end{eqnarray} and thus

\begin{eqnarray}
\frac{m'_\nu}{m_\nu} \left[ \rho_\nu - 3p_\nu \right] = \frac{\partial\rho_\nu}{\partial\phi} ~. \nonumber 
\end{eqnarray} Thus the Eq.\eqref{eqofmot} can be put in the form

\begin{eqnarray}\label{scalar}
 \ddot{\phi} +3H\dot{\phi} = -\frac{\partial \big[ U(\phi) + \rho_\nu(\phi, T_\nu) \big]}{\partial\phi}  ~.  
\end{eqnarray} indicating that the effective potential felt by the scalar field is given by

\begin{eqnarray}\label{effectivepotential1}
\mathfrak{U}(\phi, T_\nu) = U(\phi) + \rho_\nu(\phi, T_\nu) ~. 
\end{eqnarray} Of course, this potential is not uniquely defined - it can be replaced for instance by    

\begin{eqnarray}\label{effectivepotential2}
\widetilde{\mathfrak{U}}(\phi, T_\nu) = U(\phi) + \rho_\nu(\phi, T_\nu) - \rho_\nu(0, T_\nu) ~. 
\end{eqnarray} 

For the present model, the mass eigenstates of neutrinos that are nonrelativistic at present play the central role. From neutrino oscillation data the neutrino masses can be estimated in normal hierarchical spectrum as (see \cite{ParticleDataGroup:2020ssz}, page 301)

\begin{eqnarray}\label{hierarchical}
	&&m_1 \ll m_2 < m_3 ~,  \nonumber \\&& m_2 \sim 8.6\times 10^{-3}\text{\,eV}~, ~~~m_3 \sim 0.05\text{\,eV} ~.  \end{eqnarray} In view of the present CNB temperature, $T_\nu^0\simeq 1.67\times 10^{-4}$\,eV, we expect two mass eigenstates to be nonrelativistic at present since $m_2/T_\nu^0 \sim 51.5$ and $m_3/T_\nu^0 \sim 300$. For simplicity we shall consider the coupling of $\phi$ only to a single neutrino mass eigenstate. Assuming the neutrinos are nonrelativistic, $m_\nu/T_\nu \gg 1$, one can use the approximation     

\begin{eqnarray}\label{nonrelativistic}
\rho_\nu = \frac{\mathsf{g}T_\nu^4}{2\pi^2} \int_0^\infty\mathrm{d}\xi \, \frac{\xi^2\sqrt{\xi^2 +m_\nu^2/T_\nu^2}}{\mathrm{e}^\xi +1} \to \nonumber \\ \frac{3\mathsf{g}m_\nu\zeta(3)T^3_\nu}{4\pi^2}\equiv n_\nu m_\nu ~,
\end{eqnarray} where $n_\nu$ denotes the neutrino number density. Correspondingly, the Eqs.(\ref{continuity}, \ref{scalar}) simplify to

\begin{eqnarray}\label{phiindependent}
&&\dot{n}_\nu + 3Hn_\nu  =  0 ~ \Rightarrow ~ n_\nu \propto a^{-3}~ , \nonumber   \\\label{autonomous}
&&\ddot{\phi} +3H\dot{\phi} +U'(\phi) = -m'_\nu(\phi) n_\nu  ~.
\end{eqnarray} A remarkable feature of this sort of coupled model is that one can obtain an
approximate solution by minimizing the effective potential\footnote{Similar picture arises in other contexts as well \cite{Anderson:1997un, Khoury:2003rn}. }

\begin{eqnarray}\label{minimum}
U'\big(\phi_+(t)\big) + m'_\nu\big(\phi_+(t)\big) n_\nu\big(t\big) = 0 ~.
\end{eqnarray} For the validity of this approximation the slow roll conditions

\begin{eqnarray}\label{slow roll}
\dot{\phi}_+^2 \ll m_\nu(\phi_+)n_\nu~, ~~~~ |\ddot{\phi}\dot{\phi}| \ll 3H m_\nu(\phi_+)n_\nu ~, 
\end{eqnarray} are required. Namely, the Eq.\eqref{autonomous} can be put in the form

\begin{eqnarray}\label{Beimischung}
	\frac{\mathrm{d}}{\mathrm{d}t} \left(\frac{\dot{\phi}^2}{2}  + U + m_\nu n_\nu \right) = -3H \left( \dot{\phi}^2 + m_\nu n_\nu  \right) ~, 
\end{eqnarray} which immediately tells us that if the slow roll conditions \eqref{slow roll} are satisfied, then Eq.\eqref{Beimischung} is reduced to Eq.\eqref{minimum}.

Assuming we have constructed an approximate solution for a particular model, one may proceed to impose some preliminary constraints. For the equation of state parameter\footnote{Here we have omitted $p_\nu $ as neutrinos are assumed to be nonrelativistic: $p_\nu \ll \rho_\nu \sim  U(\phi_+(t_0))$.}

\begin{eqnarray}
\omega_{\text{DE}}(t_0)  = - \, \frac{U(\phi_+(t_0))}{U(\phi_+(t_0)) + \rho_\nu}  > -1 ~, \nonumber 
\end{eqnarray} one obtains the following relation

\begin{eqnarray}
	\rho^0_{DE}(1+\omega_{\text{DE}}(t_0)) \,=\, \rho_\nu^0 \,=\, n_\nu^0 m_\nu^0 ~. 
\end{eqnarray} From neutrino decoupling studies we know the ratio $n_\nu/n_\gamma$ which, by substituting the present value of CMB number density, allows one to evaluate $n^0_\nu$. Using $n_\nu^0\simeq 60$\,cm$^{-3}\simeq 10^{-10}$\,eV$^3$ (see \cite{ParticleDataGroup:2020ssz}, page 467) and taking account of the fact that at best we can take $\omega_{\text{DE}}(t_0) = - 0.99$ (see \cite{ParticleDataGroup:2020ssz}, page 495) the value of $m_2 \sim 8.6\times 10^{-3}$\, eV from Eq.\eqref{hierarchical} seems more appropriate for our discussion. This observational constraint on the equation of state parameter, $\omega_{\text{DE}}$, suggests to take

\begin{eqnarray}\label{tanapardoba}
U(\phi_+(t_0)) = 99\rho^0_\nu ~, ~~~~  \rho^0_{DE} = 100 \rho^0_\nu ~. 
\end{eqnarray}

One more observational signature that can immediately be used for constraining the model is related to the redshift at which the accelerated expansion begins. This redshift scale $<1$. Taking into account that CNB becomes nonrelativistic much earlier, one can put the condition for accelerating expansion, $\rho < -3p$, in the form

\begin{eqnarray}\label{redshift}
2U(\phi_+) > \rho_r^0(1+z)^4 + \Big(\rho_m^0 + m_\nu(\phi_+)n_\nu^0 \Big)(1+z)^3 ~. ~~~~~
\end{eqnarray} Using this relation, for a particular model one can evaluate a redshift scale at which the accelerating expansion begins.

Before discussing an approximate solution $\phi_+$ for the model under consideration, let us examine a necessary feature of the model - attractor behavior.

\section{Attractor solutions.}
\label{attractor}

After discussing the general features of the model, let us consider for the coupled model under consideration an important aspect related to the existence of attractor solution. DE model based on potential \eqref{potential} shows up an attractor behavior \cite{Ratra:1987rm} implying that irrespective to the initial values of $\phi$ and $\dot{\phi}$, the solutions tend to approach each other after a while. That is, the solutions appear to ”attract” each other in the long-time limit and, thus, any particular solution provides a good approximation to those arising from different initial conditions. That means that the model has a nice feature to avoid the fine-tuning of initial conditions\footnote{It is worth noting that the attractor behavior of the quintessence model does not mean that the question of initial conditions can be completely ignored \cite{Malquarti:2002bh, Kneller:2003xg, Liu:2004vm}.}. We need now to check whether the coupled model under consideration also has this "necessary" feature to be independent on a particular choice of initial conditions.

Since the potential \eqref{potential} satisfies the slow-rolling conditions \eqref{slow}, one may naturally expect that in the long-time limit the behavior of the system of equations  
\begin{eqnarray} && \dot{\rho}_m + 3H\rho_m  = 0 ~, \label{eins} \\ 
&& \dot{\rho}_\nu + 3H\rho_\nu  =  \frac{\beta \rho_\nu \dot{\phi}}{\phi} ~,  \label{zwei}\\ && 
\ddot{\phi} +3H\dot{\phi} +U'(\phi) = -\frac{\beta \rho_\nu}{\phi}  ~,  \label{drei}\\ && H^2 = \frac{8\pi}{3M_P^2}\left(\rho_\nu  + \rho_m  +\frac{\dot{\phi}^2}{2} + U(\phi) \right) ~, \label{vier}
\end{eqnarray} gets dominated by the potential of the scalar field. To check this, let us first assume that this is indeed the case and consider the solution of the reduced system

\begin{eqnarray}\label{asymp}
3H\dot{\phi}=\frac{\alpha V M_p^{\alpha}}{\phi^{\alpha+1}} ~, ~~ H^2 = \frac{8\pi}{3M_p^2} \frac{V M_p^\alpha}{\phi^\alpha} ~.
\end{eqnarray} The solutions that follow from Eq.\eqref{asymp} look as follows

\begin{equation}\label{asympsol}
\phi=C_1 \cdot t^{\frac{2}{\alpha+4}} ~, ~~ a=a_0 \exp\left[C_2(t^{\frac{4}{\alpha+4}}-t_0^{\frac{4}{\alpha+4}})\right] ~, 
\end{equation} where 

\begin{eqnarray}&&
C_1=\left[\frac{2\alpha V^{\frac{1}{2}} M_p^{\frac{\alpha+2}{2}}}{(\alpha+4)\sqrt{24\pi}}\right]^{\frac{2}{\alpha+4}} ~ ~\text{and} \nonumber \\&&  C_2=\frac{4}{\alpha+4} \left[ \frac{(8\pi)^{\alpha+2} V^2 (\alpha+4)}{9 (2\alpha)^{\alpha} M_p^{4}}\right]^{\frac{1}{\alpha+4}} ~. \nonumber 
\end{eqnarray} Evaluating now the asymptotic behavior of $\rho_m$ and $\rho_{\nu}$ by using the solutions \eqref{asympsol} 

\begin{eqnarray}&&
\rho_m=\rho_m^{0} \exp\left[-3C_2\left(t^{\frac{4}{\alpha+4}}-t_0^{\frac{4}{\alpha+4}}\right) \right] ~, \nonumber \\&& \rho_{\nu}=\rho_{\nu}^{0}\left(\frac{t}{t_0}\right)^{\frac{2\beta}{\alpha+4}} \exp\left[-3C_2\left(t^{\frac{4}{\alpha+4}}-t_0^{\frac{4}{\alpha+4}}\right) \right] ~, \nonumber 
\end{eqnarray} one clearly sees that $\rho_m$ and $\rho_{\nu}$ decay much faster than $U(\phi) = VM_M^\alpha/\phi^\alpha \propto t^{-2\alpha/(\alpha +4)}$. That already manifests that the Friedmann equation for $\phi_+\gtrsim M_P$ gets indeed dominated by the scalar field potential.

 A more systematic approach to the problem is based on the investigation of Eqs.(\ref{eins}, \ref{zwei}, \ref{drei}, \ref{vier}) in the phase-space. These equations of motion can be written as a first-order autonomous system

\begin{eqnarray}
&&\punkt{\zeta}_1 =   -3\zeta_1 + \sqrt{\frac{3}{2}} \zeta_4\zeta_2^2 - \nonumber  \\&& \sqrt{\frac{3}{2}}\frac{\beta}{\alpha}\zeta_4 (1 - \zeta_1^2 - \zeta_2^2 - \zeta_3^2)  +  \frac{3}{2}\zeta_1(1+\zeta_1^2-\zeta_2^2) ~. \\
&&\punkt{\zeta}_2 =   -\sqrt{\frac{3}{2}} \zeta_1\zeta_2\zeta_4  + \frac{3}{2}\zeta_2 (1+\zeta_1^2 - \zeta_2^2) ~, \\&&  \punkt{\zeta}_3 =   \frac{3}{2}\zeta_3 (\zeta_1^2 - \zeta_2^2) ~, \\&&   \punkt{\zeta}_4 =   -\frac{\sqrt{6}}{\alpha}\zeta_1\zeta_4^2 ~,
\end{eqnarray} where the dimensionless variables $\zeta_j$ are defined as
\begin{eqnarray}&&
\zeta_1 =  \sqrt{\frac{8\pi}{6}}\frac{\dot{\phi}}{M_PH} ~, ~~~ \zeta_2 = \sqrt{\frac{8\pi}{3}}\frac{\sqrt{U}}{M_PH}~,  \nonumber \\&& \zeta_3 = \sqrt{\frac{8\pi}{3}}\frac{\sqrt{\rho_m}}{M_PH} ~, ~ ~~
\label{zeta4}
\zeta_4 = -\frac{M_PU'}{\sqrt{8\pi} U} = \frac{\alpha M_P}{\sqrt{8\pi} \phi}  ~, 
\end{eqnarray} and the large over-dot denotes the derivative with respect to $N=\ln(a/a_0)$.

The asymptotic solution considered above tends to the critical point $(\zeta_1=0, \zeta_2 = 1, \zeta_3=0, \zeta_4 = 0)$. We need now to examine the stability of this critical point. For this purpose, let us first diagonalize the linear part of this system by changing the variables

\begin{eqnarray}
\zeta_1 = c_1 + \frac{c_4}{\sqrt{6}} ~, ~ \zeta_2 = 1-c_2 ~, ~ \zeta_3 = c_3 ~, ~ \zeta_4 = c_4 ~. \nonumber \end{eqnarray} As a result, one obtains

\begin{eqnarray}
	&&\punkt{c}_1 = -3c_1 + \sqrt{\frac{3}{2}}c_4 (c_2^2 - 2c_2) + \frac{c_4^2}{\alpha} \left[c_1 + \frac{c_4}{\sqrt{6}}\right]  + \nonumber \\&& \sqrt{\frac{3}{2}}\frac{\beta}{\alpha} c_4\left(2c_2 - c_2^2 - c_3^2 - \left[c_1 + \frac{c_4}{\sqrt{6}}\right]^2\right) + \nonumber \\&& \frac{3}{2} \left[c_1 + \frac{c_4}{\sqrt{6}}\right] \left(2c_2-c_2^2 + \left[c_1 + \frac{c_4}{\sqrt{6}}\right]^2\right) ~, \nonumber \\&&  \punkt{c}_2 = -3c_2 + \sqrt{\frac{3}{2}}\left[c_1 + \frac{c_4}{\sqrt{6}}\right] (1-c_2)  c_4  - \nonumber \\&&  \frac{3}{2} \left(\left[c_1 + \frac{c_4}{\sqrt{6}}\right]^2 -3c_2^2 \right) + \frac{3}{2}c_2 \left(\left[c_1 + \frac{c_4}{\sqrt{6}}\right]^2 - c_2^2  \right) ~, \nonumber \\&& \punkt{c}_3 = - \frac{3}{2}c_3 + \frac{3}{2} c_3 \left(\left[c_1 + \frac{c_4}{\sqrt{6}}\right] ^2 +2c_2 - c_2^2  \right) ~, \nonumber \\&& \punkt{c}_4 =  - \frac{\sqrt{6}}{\alpha} c_4^2 \left[c_1 + \frac{c_4}{\sqrt{6}}\right] ~. \nonumber 
\end{eqnarray} In view of the last equation, we see that the linear analysis cannot tell us whether this critical point is stable or not. Let us exploit the center manifold technique \cite{JackCarr} to reduce the system to one-dimensional invariant manifold: $c_1(c_4), c_2(c_4), c_3(c_4)$. One may define the center manifold approximately (in the vicinity of origin) by using the power series expansion of $c_j(c_4)$ functions in $c_4$. This expansion is substituted into the equations of motion $\punkt{c}_j = f_j(c_1, c_2, c_3, c_4)$, then the left-hand side is replaced by

        \begin{eqnarray}
        \punkt{c}_j = 	\frac{\mathrm{d} c_j}{\mathrm{d}c_4} \punkt{c}_4 = - \frac{\sqrt{6}}{\alpha} c_4^2 \left[c_1 + \frac{c_4}{\sqrt{6}}\right] 	\frac{\mathrm{d} c_j}{\mathrm{d}c_4} ~, \nonumber 
        \end{eqnarray} and coefficients of $c_4$ are equated. From this approach, it is clear that the leading term of $c_1(c_4)$ cannot be smaller then the second power of $c_4$. That means that in the vicinity of origin

    \begin{eqnarray}
    	\punkt{c}_4 =  - \frac{\sqrt{6}}{\alpha} c_4^2 \left[c_1 + \frac{c_4}{\sqrt{6}}\right] \approx  - \frac{c_4^3}{\alpha} ~ \Rightarrow ~ \frac{1}{2c_4^2} = \frac{N}{\alpha}  ~, \nonumber 
    \end{eqnarray} indicating that $c_4\to 0$ as $N\to\infty$. Thus, one infers that the critical point is stable and serves as an attracting fixed point.

 \section{Approximate solution $\phi_+$.}
\label{approximate}

To proceed, we need to consider the adiabaticity conditions for the approximate solution $\phi_+$ obtained by the minimization of an effective potential

   \begin{eqnarray}
   	\mathfrak{U} = V\left(\frac{M_P}{\phi}\right)^\alpha + n_\nu(t)\mu_\nu \left( \frac{\phi}{\mu_\nu} \right)^\beta ~. \nonumber 
   \end{eqnarray} One finds without much ado that

   \begin{eqnarray}
   	\phi_+(t) = \left(\frac{\alpha V M_P^{\alpha}\mu_\nu^{\beta-1}}{\beta n_\nu(t)}\right)^{\frac{1}{\alpha+\beta}} ~, ~~  \dot{\phi}_+ = \frac{3H}{\alpha+\beta}  \phi_+(t) ~. ~~~~~~\label{loesung}
   \end{eqnarray} The slow rolling conditions \eqref{slow roll} for this solution read

   \begin{eqnarray}
   	&&\label{neli1} \frac{9H^2}{(\alpha+\beta)^2} \ll  \frac{n_\nu \phi_+^{\beta-2}}{\mu_\nu^{\beta-1}} ~,  \\&&\label{neli2}  \left|\dot{H}+\frac{3H^2}{\alpha+\beta} \right| \ll n_{\nu}(t) \left(\frac{\alpha+\beta}{3}\right)^{2} \frac{\phi^{\beta-2}_+}{\mu_{\nu}^{\beta-1}} ~. 
   \end{eqnarray} Substituting

    \begin{eqnarray}
   \dot{H} = -\frac{4\pi}{M_P^2} \left( \rho_\nu + \rho_m + \dot{\phi}^2 \right) ~, \nonumber 
   \end{eqnarray} into Eq.\eqref{neli2} gives 
   
  \begin{eqnarray}\label{neli3}
  \left|\frac{8 \pi}{2 M_p^{2}} \left[ \rho_{\nu} +\rho_{m} + \frac{9H^{2}}{(\alpha+\beta)^2}\phi^2_+  \right] - \frac{3H^2}{\alpha+\beta}\right| \ll ~~~~~ \nonumber \\  n_\nu \left[\frac{\alpha+\beta}{3} \right]^2  \frac{\phi^{\beta-2}_+}{\mu_{\nu}^{\beta-1}} ~. ~~
  \end{eqnarray} Now it is plain to see that if $(\phi_+/M_P)^2 \lesssim 1$, then the Eq.\eqref{neli3} is satisfied as long as Eq.\eqref{neli1} holds. To figure out the validity condition in the present epoch for Eq.\eqref{neli1}, it is instructive to put it in the form

  \begin{eqnarray}\label{neligorva}
  \frac{9H^2\phi_+^2}{(\alpha+\beta)^2} \ll  n_\nu m_\nu ~.
  \end{eqnarray} Noting that $\alpha + \beta \sim 1$ and $H_0^2 \propto \rho_\nu^0/M_P^2$, one readily infers the validity condition of Eq.\eqref{neligorva} in the present epoch as $(\phi_+(t_0)/M_P)^2 \ll 1$. This condition that ensures the validity both of the equations (\ref{neli1}, \ref{neli2}) tells us that the coupled DE model under consideration provides the present DE in the small field regime. Let us recall from section \ref{intro} that the uncoupled model provides present DE in the large-field regime. Thus we already have one major qualitative difference that is worth paying attention.

  We shall demand that $\mu_\nu = \xi  m_\nu^0$, where $\xi \sim 1$. That is, the energy scale introduced into the model is set by the neutrino mass scale the value of which can be inferred from neutrino oscillations. Taking $\xi =1$, the solution \eqref{loesung} and the derived quantities can be written as

  \begin{eqnarray}\label{rel1}
  &&\phi_+ = m_\nu^0 \left(\frac{n_\nu^0}{n_\nu}\right)^{\frac{1}{\alpha+\beta}} ~,  \\&& \label{rel2} \frac{\dot{\phi}_+^2}{2} = \frac{9H^2(m_\nu^0)^2}{2(\alpha+\beta)^2}\left(\frac{n_\nu^0}{n}\right)^{\frac{2}{\alpha+\beta}} ~,  \\&& \label{rel3} m_\nu(\phi_+) n_\nu  = n_\nu m_\nu^0 \left(\frac{n_\nu^0}{n_\nu}\right)^{\frac{\beta}{\alpha+\beta}}  ~,  \\&& \label{rel4} U(\phi_+) =  V\left(\frac{M_P}{m_\nu^0}\right)^\alpha \left(\frac{n_\nu}{n_\nu^0}\right)^{\frac{\alpha}{\alpha+\beta}}  ~. 
  \end{eqnarray} Then from Eq.\eqref{tanapardoba} we find

  \begin{eqnarray}
  	V \,=\, 99\left(\frac{m_\nu^0}{M_P}\right)^\alpha \rho^0_\nu ~, \nonumber 
  \end{eqnarray} implying that $V\ll \rho^0_\nu$ unless $\alpha$ is taken sufficiently close to zero. From this point of view the coupled model does not look more natural then the uncoupled one. One could try, however, to use the natural reparametrization of the potential

  \begin{eqnarray}
  	U(\phi) = V\left(\frac{M_P}{\phi}\right)^\alpha  = 99\rho^0_\nu \left(\frac{m_\nu^0}{\phi}\right)^\alpha \propto \mu_\nu^4\left(\frac{\mu_\nu}{\phi}\right)^\alpha ~. \nonumber 
  \end{eqnarray}

 Using the relations (\ref{rel3}, \ref{rel4}), the Eq.\eqref{redshift} takes the form ($\rho^0_r$ is disregarded)

    \begin{eqnarray}
    	198 \left(1+z\right)^{3\alpha/(\alpha+\beta)} > \left(\frac{\rho_m^0}{\rho_\nu^0} + (1+z)^{-3\beta/(\alpha+\beta)}\right) (1+z)^3 ~. \nonumber 
    \end{eqnarray} Using a crude evaluation $\rho_{DE}\simeq 2.4\rho_m \Rightarrow\rho^0_m/\rho^0_\nu \simeq 43$ (see Eq.\eqref{tanapardoba}), one finds that in order for the beginning of accelerated expansion to obtain the redshift around $0.8$ - then one should take $3\alpha/(\alpha+\beta)\simeq 0.4 \Rightarrow\beta\simeq 7\alpha$.

   Before making any conclusions about the tree level discussion, it is necessary for internal consistency to evaluate the quantum corrections to the potential.

\section{Quantum corrections to $\mathfrak{U}$.}  
  
  The model under consideration consists of a scalar condensate that is coupled to the neutrino gas. To see the effect of quantum and thermal fluctuations, let us first evaluate the corrections coming from neutrino sector. For this propose one needs to use the technique of finite-temperature field theory \cite{Kapusta:2006pm, Bailin:1986wt} to carry out the integration over $\bar{\psi}, \psi$ in the functional integral which stands for the partition function. Performing Matsubara frequency sum, the corrections coming from neutrino sector splits into thermal and one-loop quantum corrections 
  
  \begin{eqnarray}\label{fermion}
  -  \int\frac{\mathrm{d}^3q}{(2\pi)^3} \left(\sqrt{\mathbf{q}^2+m_\nu^2(\phi)} -\sqrt{\mathbf{q}^2+\mu_\nu^2}\right) -   \nonumber \\   2T_\nu\int\frac{\mathrm{d}^3q}{(2\pi)^3} \ln \frac{1+\mathrm{e}^{-\left.\sqrt{\mathbf{q}^2+m_\nu^2(\phi)}\right/T_\nu}}{1+\mathrm{e}^{-\left.\sqrt{\mathbf{q}^2+\mu_\nu^2}\right/T_\nu}}   ~. 
  \end{eqnarray} We have chosen the normalization ensuring that the contribution vanishes when there is no coupling: $\beta = 0 \Rightarrow m_\nu = \mu_\nu$. At the same time it means that this contribution vanishes for $t=t_0$ as we have assumed in previous section $m^0_\nu = \mu_\nu$. The second term in Eq.\eqref{fermion} is convergent, while the first integral needs some sort of regularization to render the
  divergent momentum integration finite. Treating the first integral in a standard manner, one may first put it in a Wick rotated $4$-dimensional form \cite{Kapusta:2006pm, Bailin:1986wt} and then exploit a $4$-momentum cutoff or one can immediately use a 3-momentum cutoff. In both cases the result contains quartic, quadratic and logarithmic terms in cutoff. The model under consideration is not renormalizable and therefore we cannot use a standard renormalization procedure \cite{Coleman:1973jx, Weinberg:1973am} to absorb divergent terms into the bare parameters of the theory. There are, however, certain arguments based on Lorentz symmetry and the application of different regularization schemes for evaluating Nullpunktsenergie in field theory showing that while the logarithmic divergence is universal, the power-law divergences are not \cite{Akhmedov:2002ts, Ossola:2003ku, Koksma:2011cq}. There is one more approach based on the method of $\zeta$-function regularization \cite{Ramond:1981pw} that immediately gives the logarithmic term which contains the renormalization scale $\mu$ but no regularization parameter. Using this result quantum corrections coming from Eq.\eqref{fermion} can be evaluated as

  \begin{eqnarray}\label{qcfermion}
  	- \frac{m_\nu^4(\phi)}{32\pi^2}\ln \frac{m_\nu^2(\phi)}{\mu^2}  +  \frac{\mu_\nu^4}{32\pi^2}\ln \frac{\mu_\nu^2}{\mu^2} ~. 
  \end{eqnarray}  
  
  The second integral in Eq.\eqref{fermion} can be expressed simply in low and high-temperature limits \cite{Kapusta:2006pm, Bailin:1986wt}. What we need now is the low-temperature approximation

  \begin{eqnarray}
  2T_\nu\left[ \left(\frac{m_\nu T_\nu}{2\pi}\right)^{3/2}\mathrm{e}^{-m_\nu/T_\nu} - \left(\frac{\mu_\nu T_\nu}{2\pi}\right)^{3/2}\mathrm{e}^{-\mu_\nu/T_\nu}  \right]  ~. ~~~~~~
  \end{eqnarray} In the low-temperature regime, this term can be neglected as compared to $n_\nu m_\nu$ in $\mathfrak{U}$. It can be readily verified by recalling that (see Eq.\eqref{nonrelativistic}) $n_\nu m_\nu \propto T_\nu^3m_\nu$.

   Apart from the quantum corrections arising due to coupling with CNB, there is an additional contribution that comes due to quantum fluctuations of $\phi$. This contribution can be evaluated by the Nullpunktsenergie

   \begin{eqnarray}\label{Nullpunktsenergie}
   \frac{1}{2}\int\frac{\mathrm{d}^3 q}{(2\pi)^3} \left[\sqrt{\mathbf{q}^2 + m_{\text{eff}}^2(\phi)} -q \right] ~, 
   \end{eqnarray} where

   \begin{eqnarray}
   &&m_{\text{eff}}^2(\phi) \equiv  \mathfrak{U}''(\phi) =  \nonumber \\&& ~~~~ \frac{\alpha(\alpha+1)VM_P^\alpha}{\phi^{\alpha+2}}  + \frac{n_\nu\beta(\beta-1)}{\mu_\nu^{\beta-1}} \phi^{\beta-2}  ~, \label{effectivemass}
   \end{eqnarray} and by subtracting $q$ from the integrand in Eq.\eqref{Nullpunktsenergie} we have chosen a normalization that the integral vanishes when $m_{\text{eff}}^2(\phi) = 0$. The reason we have evaluated the effective mass of $\phi$ by the tree-level effective potential $\mathfrak{U}$ (without including in it quantum corrections coming from the coupling with neutrinos) is that all of the quantum corrections we are discussing now are of the order of $\hbar$. The integral \eqref{Nullpunktsenergie} can be evaluated in much the same way as the first integral in Eq.\eqref{fermion}. Therefore, we arrive at the expression similar to \eqref{qcfermion}

   \begin{eqnarray}\label{qcphi}
   \frac{m_{\text{eff}}^4(\phi)}{64\pi^2}\ln \frac{m_{\text{eff}}^2(\phi)}{\mu^2} ~. 
   \end{eqnarray}

   Now let us see how does the net effect of quantum corrections alter the dynamics of $\phi$. First by assuming $\beta > 2$ one infers from Eq.\eqref{effectivemass} that $m_{\text{eff}}^4(\phi)$ has a minimum at

   \begin{eqnarray}
   	\phi = \left(\frac{\alpha(\alpha+1)(\alpha
   		+2)VM_P^\alpha \mu_\nu^{\beta-1}}{\beta(\beta -1)(\beta-2)n_\nu}\right)^{1/(\alpha+\beta)} ~ , \nonumber 
   \end{eqnarray} that is quite close to $\phi_+$ given by the Eq.\eqref{loesung}. Thus, by adding the expression \eqref{qcphi} to the tree-level effective potential the physical picture remains pretty much the same. The situation is somewhat different with the contribution \eqref{qcfermion}. Setting $\mu = \mu_\nu$, the quantum corrected effective potential takes the form 
   
   \begin{eqnarray}
   	&&V\left(\frac{M_P}{\phi}\right)^\alpha + n_\nu \mu_\nu \left(\frac{\phi}{\mu_\nu}\right)^\beta + \nonumber \\&& ~~~	\frac{m_{\text{eff}}^4(\phi)}{64\pi^2} \ln\frac{m_{\text{eff}}^2(\phi)}{\mu_\nu^2}  -  \frac{m_\nu^4(\phi)}{32\pi^2}\ln\frac{m_\nu^2(\phi)}{\mu_\nu^2}  ~, \nonumber 
   \end{eqnarray} where it is understood that $m_{\text{eff}} >\mu_\nu$ since Eq.\eqref{Nullpunktsenergie} is positive definite. We see that quantum correction coming from CNB is positive in the past (it is in accordance with Eq.\eqref{fermion}) but in the future it becomes negative. One more significant fact is that in the past this term is small as compared to $n_\mu\nu_\nu(\phi/\mu_\nu)^\beta$ and therefore does not affect the minimum of the effective potential in any significant way. However, in the near future the absolute value of this term becomes greater than $n_\mu\nu_\nu(\phi/\mu_\nu)^\beta$ and the effective potential cannot develop any more the minimum. To judge about the behavior of the potential for large values of $\phi$ - one needs to consider two-loop and higher corrections to the effective potential.

\section{Discussion.}
\label{discussion}

In view of the preceding discussion, let us figure out what might be called beneficial features of the coupling of inverse power-law potential with the neutrinos. One of the qualitative differences between coupled and uncoupled models is that in the coupled model the DE is provided in the small field regime: $\phi \simeq \mu_\nu$ that maybe considered as a good feature for saving the model against the quantum corrections. However, in itself the question of evaluating the size of one-loop corrections in models that are non-renormalizable is highly nontrivial. Taking all the cutoff dependent terms (quartic and quadratic) as the valid result for the corrections maybe really troublesome as the theory completely lacks its predictive power \cite{Kolda:1998wq, Doran:2002bc}. There is however some hope to avoid the quartic and quadratic terms in cutoff by demanding the Lorentz symmetry \cite{Akhmedov:2002ts, Ossola:2003ku, Koksma:2011cq} or alternatively one can use the $\zeta$-function regularization for obtaining a cutoff independent result \cite{Ramond:1981pw}. As we have seen even the logarithmic terms that come from quantum corrections can qualitatively alter the model in the future. However, these terms are harmless up to the present. In the end, let us comment about the obvious difference between coupled and uncoupled models in light of the quantum corrections due to graviton loops. As long as we are in the small field regime, we can safely ignore the corrections coming from the one-loop graphs involving gravitons since they are expressed in terms of the quantities \cite{Smolin:1979ca}

\begin{eqnarray}\label{graviton}
\frac{\mathfrak{U}(\phi)}{M_P^4} ~, ~~ \frac{\mathfrak{U}'(\phi)}{M_P^3}  ~,~~ \frac{\mathfrak{U}''(\phi)}{M_P^2} ~. \nonumber 
\end{eqnarray} They may of course not be suppressed in the large field regime $\phi\gtrsim M_P$, which is important for providing DE in the uncoupled model.

Both coupled and uncoupled models show up the attractor behavior and there is no difference from this point of view. However, the coupled model provides a convenient way to find an approximate solution for $\phi$ by minimizing the effective potential. In the case of uncoupled model, one needs to specify the dominant energy component in the Friedmann equation for finding an analytic solution for $\phi$. In addition an approximate solution in the coupled model has a nice feature that it is written as an explicit function of $n_\nu = n_\nu^0(1+z)^{3}$. It allows one to parameterize the quantities $\dot{\phi}^2_+/2,\, U(\phi_+),\, m_\nu(\phi_+)n_\nu$ in terms of the redshift parameter (see Eqs.(\eqref{rel1} - \eqref{rel4}) and put down the Friedmann equation in the form

	\begin{eqnarray}
	H^2 = \frac{8\pi}{3M_P^2} \left(\frac{9H^2(m_\nu^0)^2}{2(\alpha+\beta)^2}\left(1+z\right)^{\frac{-6}{\alpha+\beta}}  +    \rho^0_m(1+z)^3 +  \right. \nonumber \\ \left.m_\nu^0 (n_\nu^0)^{\frac{\alpha +   2\beta}{\alpha+\beta}} (1+z)^{\frac{3\alpha}{\alpha+\beta}} + V\left(\frac{M_P}{m_\nu^0}\right)^\alpha \left(1+z\right)^{\frac{3\alpha}{\alpha+\beta}} \right) ~, \nonumber 
	\end{eqnarray} that can be readily used for determining the best-fit parameters of model in view of the low-redshift observations such as Hubble parameter and BAO distance measurements. There is however a subtle point related to the fact that in most cases the mass varying neutrino models predict the clumpy structure of CNB \cite{Afshordi:2005ym}. The nucleation of neutrino nuggets takes place when neutrinos enter the non-relativist regime and one needs to work out the back-reaction effect of these nuggets on the background cosmology \cite{Schrempp:2009kn, Nunes:2011mw} for considering low-redshift observational constraints.

Finally, let us look at the parameter fine-tuning problem. The DE density is $\propto \mu_\nu^4$. In the uncoupled model the onset of DE takes place for $\phi\sim M_P$. Parameterizing the potential as $V^{4+\alpha}/\phi^\alpha$, one should require that $V^{4+\alpha}\sim \mu_\nu^4M_P^\alpha$. We have three different mass scales but the real situation is that the theory predicts the scale $M_P$ and then we need to adjust $V$ in such a way as to reproduce the mass scale $\mu_\nu$. The coupled model is naturally parameterized with a single parameter $\mu_\nu$, that is, $VM_P^\alpha$ is set to be $\propto \mu_\nu^{4+\alpha}$. The question now is to ask what predictions are provided by the model. In view of the Eq.\eqref{rel4} we see that $U(\phi_+)$, which provides the main contribution to the $\rho_{DE}$ (see Eq.\eqref{tanapardoba}), behaves as 
\begin{eqnarray}
	\mu_\nu^4\left(\frac{T_\nu}{\mu_\nu}\right)^{-3\alpha/(\alpha+\beta)} = \mu_\nu^4\left(\frac{T_\nu^0}{\mu_\nu}\right)^{-3\alpha/(\alpha+\beta)}  a^{-3\alpha/(\alpha+\beta)} ~, \nonumber 
\end{eqnarray} and irrespective to the precise values of $\alpha$ and $\beta$ ensures that in the future it becomes dominant as compared to matter whose energy density decays as $\rho_m^0a^{-3}$. Since $\rho_m^0 \sim \mu_\nu^4$, the takeover occurs when $T_\nu \sim \mu_\nu$. Does it mean that the temperature-scale at which neutrinos become nonrelativistic can appear in model dynamically? One may hope that the motion of scalar field becomes damped when $T_\nu \sim \mu_\nu$ by the term $\rho_\nu-3p_\nu$ that enters Eqs.(\ref{continuity}, \ref{eqofmot}) since one may naturally expect that this term is small when neutrinos are relativistic and becomes significant only when CNB cools down. It would be a nice feature of the model if the cooling of CNB to the temperature $\sim \mu_\nu$ really triggers some dynamical effect that could explain the activation of DE. However, nothing similar happens in the model. Namely, for "hot" CNB, the term $T_\nu^3m_\nu$ in the effective potential is replaced by $T_\nu^2m_\nu^2$ and the mnimum of the effective potential occurs at 

\begin{eqnarray}
	\phi_+ \sim \left(\frac{\alpha\mu^{2+\alpha+2\beta}}{\beta T_\nu^2}\right)^{1/(\alpha+2\beta)}  \Rightarrow~ \dot{\phi}_+ = \frac{\phi_+2H}{\alpha+2\beta} ~. 
\end{eqnarray} In deriving this equation we have tacitly assumed that $VM_P^\alpha \propto \mu_\nu^{4+\alpha}$. In the case of "hot" CNB, the Eq.\eqref{Beimischung} gets replaced by (it is just a standard continuity equation that follows from Eqs.(\ref{continuity}, \ref{eqofmot})) 

\begin{eqnarray}\label{phiplus}
\frac{\mathrm{d}}{\mathrm{d}t} \left(\frac{\dot{\phi}^2}{2} + U + \rho_\nu\right) = -3H\left(\dot{\phi}^2+ \rho_\nu + p_\nu\right) ~, \nonumber 
\end{eqnarray} and the slow-rolling conditions \eqref{slow roll} read now

\begin{eqnarray}
	\dot{\phi}^2_+ \ll \rho_\nu ~, ~~ |\dot{\phi}_+\ddot{\phi}_+| \ll  3H \rho_\nu ~. \nonumber 
\end{eqnarray} As in the case of cooled CNB, it is straightforward to verify that for $T_\nu$ by a few orders of magnitude greater than $\mu_\nu$, slow rolling conditions are again satisfied with a high precision because of the suppression factor $(\phi_+/M_P)^2$ that appears in $\dot{\phi}^2_+$ and $\dot{\phi}_+\ddot{\phi}_+$. Thus, the time-scale at which CNB enters the nonrelativistic regime does not play any special dynamical role in the model. The field starts to enter the slow rolling regime much before this time. It maybe of interest to note that if in the uncoupled model the region for initial values of field for providing DE is $0 <\phi \lesssim M_P$, in the case of coupled model this region becomes smaller $0 <\phi \lesssim \mu_\nu$.

As it is discussed in Section \ref{attractor}, for large values of $\phi$, the model gets practically uncoupled and the motion of field is determined by the inverse power-law potential solely. That means that in the remote future the model will again enter the phase of DE dominance.

\begin{acknowledgments} 
	The useful discussions with Gennady Chitov and Bharat Ratra are kindly acknowledged. The work of M. M. was supported in part by the Rustaveli National Science Foundation of Georgia under Grant No. FR-19-8306. The work of V. T. was supported in part by the Rustaveli National Science Foundation of Georgia under Grant No. PHDF-22-3918. 
\end{acknowledgments}

\nocite{*}
\bibliography{References}

\end{document}